\documentclass[aps,amsfonts,epsf]{revtex4}
\usepackage{graphics}
\usepackage{graphicx}
\usepackage{epsfig}

\begin{document}
\input epsf.tex

\title{Modelling the post-Newtonian test-mass gravitational wave flux function for compact binary systems using Chebyshev polynomials.}
\author{Edward K. Porter}

\affiliation{Dept. of Physics, Montana State University, Bozeman, 59717, MT, USA.} 

\vspace{1cm}
\begin{abstract}
\noindent We introduce a new method for modelling the gravitational wave flux function of a test-mass particle inspiralling into an intermediate mass Schwarzschild black hole which is based on Chebyshev polynomials of the first kind.  It is believed that these Intermediate Mass Ratio Inspiral events (IMRI) are expected to be seen in both the ground and space based detectors.  Starting with the post-Newtonian expansion from Black Hole Perturbation Theory,  we introduce a new Chebyshev approximation to the flux function, which due to a process called Chebyshev economization gives a model with faster convergence than either post-Newtonian or Pad\'e based methods.  As well as having excellent convergence properties, these polynomials are also very closely related to the elusive minimax polynomial.    We find that at the last stable orbit, the error between the Chebyshev approximation and a numerically calculated flux is reduced, $< 1.8\%$, at all orders of approximation.  We also find that the templates constructed using the Chebyshev approximation give better fitting factors, in general $> 0.99$, and smaller errors, $< 1/10\%$, in the estimation of the Chirp mass when compared to a fiducial exact waveform, constructed using the numerical flux and the exact expression for the orbital energy function, again at all orders of approximation.  We also show that in the intermediate test-mass case, the new Chebyshev template is superior to both PN and Pad\'e approximant templates, especially at lower orders of approximation.
\end{abstract}

\maketitle

\section{Introduction}
The inspiral of compact objects such as neutron stars and black holes are expected to be a major source of gravitational waves (GW) for the ground-based detectors LIGO, VIRGO, GEO600 and TAMA~\cite{LIGO, VIRGO, GEO, TAMA}, as well as for the future planned space-based detector LISA~\cite{LISA}.  Due to the effect of radiation reaction, the orbit of the binary system slowly decays over time.  As this happens the amplitude and frequency of the waveform increases emitting a `chirp' waveform.

There have been many efforts to create templates which will approximate a possible signal to high accuracy.  On one hand we have the post-Newtonian (PN) expansion of Einstein's equations to treat the dynamics of the system~\cite{BDIWW,BDI,WillWise,BIWW,Blan1,DJSABF,BFIJ}.  This works well in the adiabatic or slow-motion approximation for all mass ranges.  On the other hand we have black hole perturbation theory~\cite{Poisson1,Cutetal1,TagNak,Sasaki,TagSas,TTS} which works for any velocity, but only in situations where the mass of one body is much greater than the other. While templates have been generated to 5.5 PN order for a test-mass orbiting a Schwarzschild black hole~\cite{TTS}, and to 3.5 PN order for non-spinning binaries of comparable mass~\cite{DJSABF,BFIJ}, a number of difficulties still need to be tackled.  The main problem is that both templates are a function of the orbital energy and GW flux functions.  In the test-mass case, an exact expression in known for the orbital energy, but we have a PN expansion for the flux function.  In the comparable-mass case, a PN expansion is known for both functions.  It has been shown that the convergence of both methods is too slow to be useful in creating templates that can be confidently used in a GW search~\cite{TTS,Cutetal2,Poisson3,Poisson4,DIS1}.  We also know that the PN approximation begins to break down when the orbital separation of the two bodies is $r\leq 10 M$~\cite{Brady}.  This means that as we approach the Last Stable Orbit (LSO) the templates begin to go out of phase with a possible signal due to the increase of relativistic effects.  As most search methods are based on matched filtering, any mismatch in phase between our templates and a signal will result in a loss of recovered signal-to-noise ratio (SNR) and an increase in the error in the estimation of parameters.

It was shown~\cite{DIS1, portersathya, porter2} that templates based on resummation methods such as Pad\'e approximation have a faster convergence in modelling the gravitational waveform.  The Pad\'e based templates were then used to partially construct Effective One Body templates~\cite{BD, Damour01} which went beyond the adiabatic approximation and modelled the waveform into the merger phase.  Other more phenomenological templates such as the BCV~\cite{BCV1, BCV2, BCV3} templates seem to be excellent at detecting GW, but are not necessarily the best template to use in the extraction of parameters.  

In this paper we focus on the inspiral of IMRI sources.  These sources encompass the inspiral of a Neutron star (NS) into a black hole with masses of 10s to 100s of solar masses (which should be observable in the ground based detectors), to the inspiral of a low mass Supermassive black hole ($10^{3}-10^{4}M_{\odot}$) into a more massive black hole ($10^{6}-10^{9}M_{\odot}$) as should be observable with future space based detectors.  We don't believe that the PN approximation used here will be sufficiently accurate to model EMRI sources, and expect that other methods such as analytic and numerical kludge waveforms~\cite{CutlerBarack, K1, K2, K3} will be used to properly model EMRI waveforms.  On the other hand, we fully believe that this method of resumming the PN series using Chebyshev polynomials will be also applicable to the comparable mass case~\cite{porter3}.

Throughout the paper we use the units $c = G = 1$.

\subsection{Improving Template Construction Using the Minimax and Chebyshev Polynomials.}
The problem with expansions like a Taylor series is that they are based on Weierstrass's theorem, which assumes that there are enough terms in the expansion to sufficiently model the function we are approximating.  We know from previous studies that the 11 term expansion for the flux function for test-mass systems may not be sufficient.  A more promising possibility is based on getting close to the minimax polynomial by using the family of Ultraspherical (or Gegenbauer) polynomials which are defined by
\begin{equation}
P_{n}^{(\alpha)}(x)=C_{n}\left(1-x^{2}\right)^{-\alpha}\frac{d^{n}}{dx^{n}}\left(1-x^{2}\right)^{n+\alpha}\,\,\,\,\,\,\,\,\,\,\,\,\,\left(-1\leq\alpha\leq\infty\right),
\end{equation}
where $C_{n}$ is a constant.  These polynomials are orthogonal over $x\in[-1,1]$ with respect to the weight function $\left(1-x^{2}\right)^{\alpha}$.  A feature of the polynomials $P_{n}^{(\alpha)}(x)$ is that they have $n$ distinct and real zeros and exhibit an oscillatory behaviour in the interval $[-1,1]$.  For $\alpha=-1/2$ the amplitude of the oscillations remain constant throughout the interval and is conducive to trying to find an "equal-ripple" error curve, which is integral to the minimax polynomial.

This value of $\alpha$ corresponds to the Chebyshev polynomials of the first kind, $T_{n}(x)$, (hereafter Chebyshev polynomials).  These polynomials are closely related to the minimax polynomial due to the fact that there are $n+1$ points in [-1,1] where $T_{n}(x)$ attains a maximum absolute value with alternating signs, i.e. $|T_{n}(x)|=\pm 1$~\cite{Mason}.  It can be shown~\cite{Snyder} that the Chebyshev polynomials exhibit the fastest convergence properties of all of the Ultraspherical polynomials.  

For our purposes, we need to approximate polynomials which are a function of the dimensionless velocity $v$ in the domain $v\in[0, v_{lso}]$, where $ v_{lso}=1/\sqrt{6}$ is the velocity at the LSO for a test-particle orbiting a Schwarzschild black hole.  In this case we use the Shifted Chebyshev polynomials, designated $T_{n}^{*}(v)$.  We can also transform from the interval $[1,-1]$ to an arbitrary interval $[a,b]$ using
\begin{equation}
s=\frac{2x-(a+b)}{b-a}\,\,\,\,\,\,\,\,\,\,\,\,\,\,\,\ \forall\,\, x\in[a,b], s\in[-1,1].
\end{equation}
In this case we have
\begin{equation}
s =  \frac{2v}{v_{lso}}-1 = \sqrt{24}v-1\;\;\;\;\;\;\forall\,\, v\in[0,v_{lso}].
\end{equation}
We can now write the shifted Chebyshev polynomials in the form
\begin{equation}
T_{n}^{*}(v)=T_{n}(s)=T_{n}(\sqrt{24}v-1),
\end{equation}
and the recurrence relation as 
\begin{equation}\label{eqn:rec}
T_{n}^{*}(v)=2(\sqrt{24}v-1)T_{n-1}^{*}(v)-T_{n-2}^{*}(v),
\end{equation}
such that the shifted polynomials have the initial conditions
\begin{equation}\label{eqn:init}
T_{0}^{*}(v)=1 ,\,\,\,\,\,\,\,\,\,\,T_{1}^{*}(v) = \sqrt{24}v-1.
\end{equation}

\section{The Test-Mass Gravitational Waveform.}\label{sec:waveform}
In the stationary phase approximation the Fourier transform for positive frequencies is given by~\cite{Thorne,SathDhur,DWS,DIS2}
\begin{equation}
\tilde{h}(f) \equiv \int_{-\infty}^\infty h(t) \exp(2\pi i f t)\, dt={\mathcal A}f^{-7/6}
e^{i\left[\psi(f)-\frac{\pi}{4}\right]},
\label{d4.6a}
\end{equation}
where ${\mathcal A}$ is a normalization constant.  The phase of the Fourier transform in the stationary phase approximation, $\psi(f)$, is found by solving the set of $1^{st}$ order ODEs given by
\begin{equation}
\frac{d\psi}{df} - 2\pi t = 0, \ \ \ \
\frac{dt}{df} + \frac{\pi m^2}{3v_f^2} \frac{E'(f)}{ F(f)} = 0,
\label {eq:frequency-domain ode}
\end{equation}
where $m$ is the total mass, $v_f = (\pi m f)^{1/3}$ is the instantaneous velocity, $E'(v)=dE/dv$ is the derivative of the orbital energy with respect to the velocity $v = (m\Omega)^{1/3} = x^{1/2}$, where $\Omega$ is the angular velocity as observed at infinity and $x$ is an invariant velocity parameter observed at infinity. Finally, $F(v)$ is the gravitational wave flux function. 

For a test-mass particle in circular equatorial orbit about a Schwarzschild black hole, an exact expression for the orbital energy exists~\cite{Chandra}.  Its derivative with respect to the velocity is given by
\begin{equation}
E'(v) = -\eta v\frac{1-6v^2}{\left(1-3v^2\right)^{3/2}},
\end{equation}
where we have introduced a finite-mass dependence through the reduced mass ratio, $\eta=m_{1}m_{2}/m^{2}$. From this equation the LSO is found by demanding $E'(v) = 0$, giving $v_{lso}=1/\sqrt{6}$.  For the flux function we only have a PN expansion of the form
\begin{equation}
F_{T_{n}}(v) = F_{N}(v)\left[\sum_{k=0}^{11}\,a_{k}v^{k} + 
\ln(x)\,\sum_{k=6}^{11}\,b_{k} v^{k} + {\mathcal O}\left(v^{12}\right)\right],
\label{eq:flux}
\end{equation}
where $F_{N}(x)$ is the dominant {\it Newtonian} flux function given by
\begin{equation}
F_{N}(x) = \frac{32}{5}\eta^{2}v^{10},
\end{equation}
and the coefficients in the expansion of the flux function are given by~\cite{Poisson1,Cutetal1,TagNak,Sasaki,TagSas,TTS}.  We begin to encounter logarithmic terms at $k=6$ and above.  It is well know that terms such as these can destroy the convergence of a power series expansion.

The Pad\'e approximant to the flux is defined as
\begin{equation}
F_{P_{n}}(v) = \left(1-\frac{v}{v_{pole}} \right)^{-1}F_{N}\left[1+\ln\left(\frac{v}{v_{lso}}\right)\sum_{k=6}^{11}\,l_{_{k}}v^{k}\right]
P^{N}_{M}\left[\sum_{k=0}^{11}\,f_{_{k}}v^{k}\right],
\end{equation}
where $P_{M}^{N}$ is the Pad\'e operator, $v_{pole} = 1/\sqrt{3}$ is the velocity at the photon ring and the coefficients $l_{k}$ and $f_{k}$ are related to the coefficients in the original PN expansion.

\section{The intermediate test-mass Chebyshev approximation flux function to 5.5-PN order.}\label{sec:ChebApprox}
After factoring out the logarithmic term and introduce a linear term into the non-logarithmic series,  the next step is to expand both power series in the above equation in terms of the shifted Chebyshev polynomials.  This is done by writing each monomial in the power series in terms of the shifted Chebyshev polynomials and substituting back into the series above.  So, for example, starting with Equations~(\ref{eqn:rec}) and (\ref{eqn:init}), we can invert each expression for the monomials in $v$, i.e.
\begin{equation}
1 = T^{*}_{0}(v),\,\,\,\,v = (24)^{-1/2}\left[T^{*}_{0}(v)+T^{*}_{1}(v)\right]...
\end{equation}
and so on.

Proceeding like this for all monomials, it then allows us to write the power series in the PN expansion solely in terms of the shifted Chebyshev polynomials.  The first advantage the Chebyshev approximation has over both the PN and Pad\'e approximations is that we can also expand the power series appearing in the logarithmic terms as a Chebyshev series.  Substituting for the monomials, we can write
\begin{equation}
\sum_{k=6}^{11}\,l_{_{k}}v^{k}=\sum_{k=0}^{11}\,\xi_{_{k}}T_{n}^{*}(v),\,\,\,\sum_{k=0}^{11}\,f_{_{k}}v^{k}=\sum_{k=0}^{11}\,\lambda_{_{k}}T_{n}^{*}(v).
\end{equation}
As the expression for both series in terms of the coefficients $\xi_{_{k}}$, $\lambda_{_{k}}$ are long so we will omit them here. 
We should emphasise the fact here that although the values of the coefficients $l_{k}$ are zero up to $k=6$, the Chebyshev expansion includes terms from $k=0$.  This allows us to define the Chebyshev approximation to the gravitational wave flux function as
\begin{equation}
F_{C_{n}}(v) = \left(1-\frac{v}{v_{pole}}\right)^{-1}F_{N}(v)\left[1+\ln\left(\frac{v}{v_{lso}}\right)\sum_{k=0}^{11}\,\xi_{_{k}}T_{n}^{*}(v)\right]
\left[\sum_{k=0}^{11}\,\lambda_{_{k}}T_{n}^{*}(v)\right],
\end{equation}
where we re-introduce the pole at the photon ring.
\begin{table}\label{tab:truncerr}
\begin{tabular}{c|c|c|c|c|c|c|c}\hline
$n$ & $5$ & $6$ & $7$ & $8$ & $9$ & $10$ & $11$ \\ \hline 
$\epsilon_{_{T_{n}}}$ & $4.3\times 10^{-1}$ & $4.7\times10^{-1}$ & $7.8\times10^{-2}$ & $5.4\times10^{-2}$ & $1.7\times10^{-1}$ & $1.4\times10^{-1}$ & $2.8\times10^{-1}$ \\
$\epsilon_{_{C_{n}}}$ & $9.9\times 10^{-4}$ & $1.8\times10^{-4}$ & $7.6\times10^{-5}$ & $2.6\times10^{-5}$ & $5\times10^{-6}$ & $8.5\times10^{-7}$ & $7.1\times10^{-8}$ \\ 
 & & & & & & & \\ \hline
$\epsilon_{_{T_{n}}}^{tot}$ & $1.37$ & $0.94$ & $0.47$ & $0.39$ & $0.34$ & $0.17$ & $2.8\times10^{-1}$ \\
$\epsilon_{_{C_{n}}}^{tot}$ & $1.3\times 10^{-3}$ & $2.9\times10^{-4}$ & $1.1\times10^{-4}$ & $3.2\times10^{-5}$ & $5.9\times10^{-6}$ & $9.2\times10^{-7}$ & $7.1\times10^{-8}$ \\ 
 & & & & & & & \\ \hline
\end{tabular}
\caption{The top two lines give the truncation errors associated with each order of approximation for both the PN and Chebyshev flux functions at the LSO, where the error in the approximation is greatest.  The bottom two lines give the total truncation error incurred as we reduce the order of approximation from 5.5 to 2 PN.}
\end{table}

\subsection{Modelling the test-mass flux function.}\label{sec:model}
A major reason for using Chebyshev polynomials is a process called Chebyshev economization~\cite{Lanczos}.  This works as follows : we can always expand a power series $p_{n}(x)$ of some function $f(x)$ in terms of a Chebyshev series $q_{n}(x)$ such that
\begin{equation}
p_{n}(x) =\sum_{k=0}^{n}a_{k}x^{k} = \sum_{k=0}^{n}b_{k}T_{k}(x) = q_{n}(x).
\end{equation}
In general the series $p_{n}(x)$ diverges as we move away from the point of expansion.  Also, we usually find that $|a_{n}|>|a_{n-1}|$.  Therefore if we wish to truncate the polynomial $p_{n}(x)$ to give another polynomial $r_{m}(x)$, such that $m<n$, we introduce an additional truncation error of the order $|a_{n}x^{n}|$ in $r_{m}(x)$.  On the other hand, the coefficients $b_{k}$ in a Chebyshev expansion always decrease as we increase the order of approximation such that $|b_{n}|<|b_{n-1}|$.  The same truncation with the Chebyshev series $q_{n}(x)$ to give another Chebyshev series $s_{m}(x)$, gives an error which is guaranteed to be smaller than an equivalent truncation in the Taylor series.  This is due to the fact that as $|T_{n}(x)|\leq 1$ in the interval,  the truncation error is of the order of the truncated coefficient $|b_{n}|$ which is in general a small number.  The other main advantage of using Chebyshev polynomials is the fact that because they use information from the entire interval, their convergence has absolutely nothing to do with the convergence of the initial Taylor series.  

It has been shown in the past that for a test-mass in a circular equatorial orbit around a Schwarzschild black hole, that resumming the PN flux approximation using Pad\'e approximation produces a better fit to the numerical flux.  However, this method does suffer from certain aspects such as poles and singularities, and the fact that it may not perform as well as the PN flux at low orders of approximation.  In this section we investigate the modelling of the numerical flux~\cite{Shibata}, calculated using the Teukolsky formulism from black hole perturbation~\cite{BHPT}, using Chebyshev economization.  It has also been well known that the PN approximation has its largest error at the LSO.   We would expect as $v$ takes on a maximum value at the LSO, that the maximum truncation error should be well approximated by the first neglected term, i.e. for $F_{T_{n-1}}(x)$ the error should be smaller than
\begin{equation}
\epsilon_{T_{n}} = a_{n}v_{lso}^{n}+\ln(v_{lso}) b_{n}v_{lso}^{n}.
\end{equation}
For the Chebyshev approximation to the flux, we know that over the entire interval $v\in[0,v_{lso}]$ the shifted Chebyshev polynomials have a maximum absolute value of $|T_{n}^{*}(v)|\leq \pm1$.  Therefore, the induced truncation error will be well approximated by the size of the coefficient of the truncated term, i.e.
\begin{equation}
\delta_{C_{n}} \leq \lambda_{n}, 
\end{equation}
due to the fact that at the LSO, the logarithmic terms are killed off.
\begin{figure}[t]
\vspace{0.25 in}
\centerline{\hbox{ \hspace{0.0in} 
    \epsfxsize=2.5in
    \epsffile{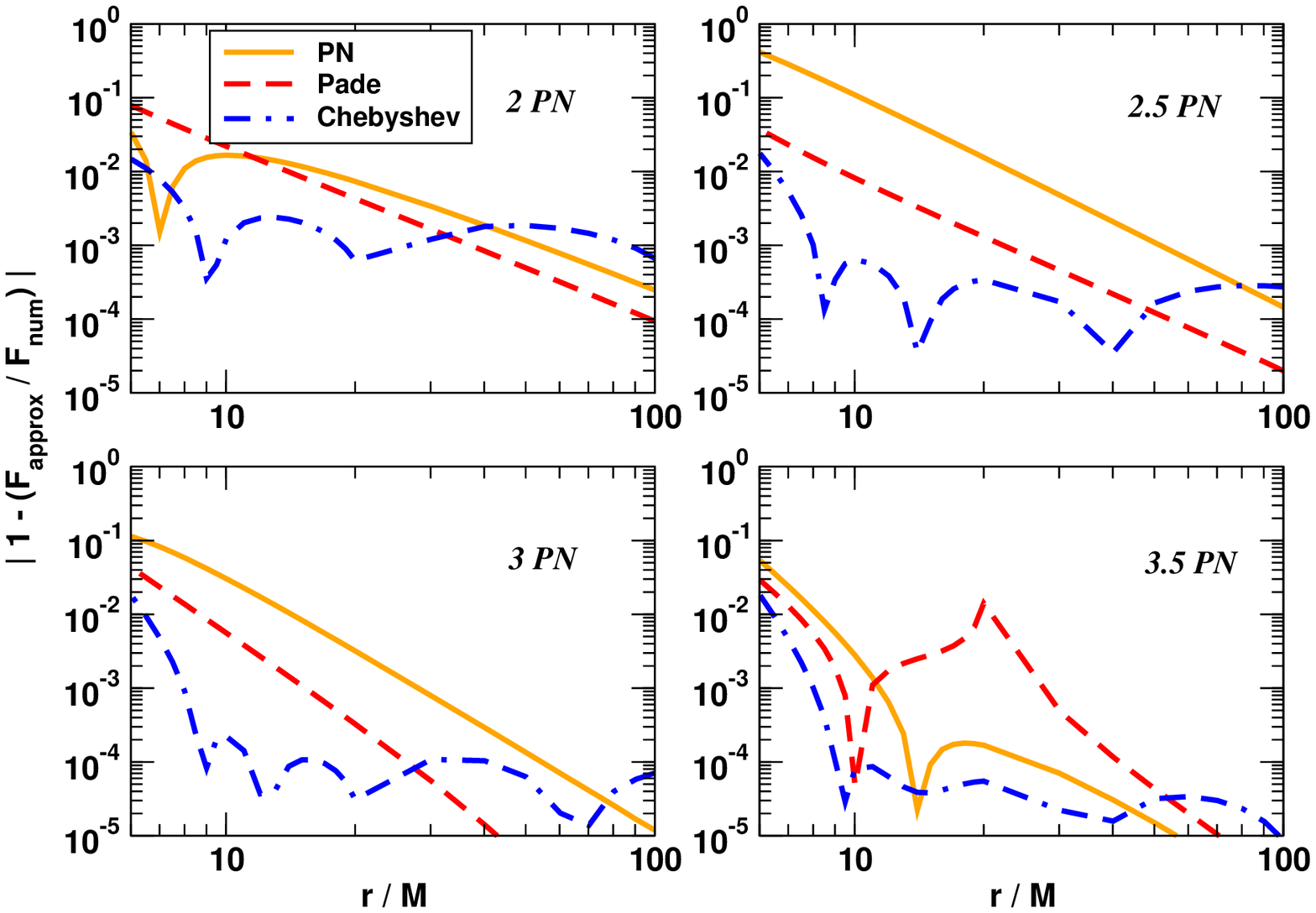}
    \hspace{0.25in}
    \epsfxsize=2.5in
    \epsffile{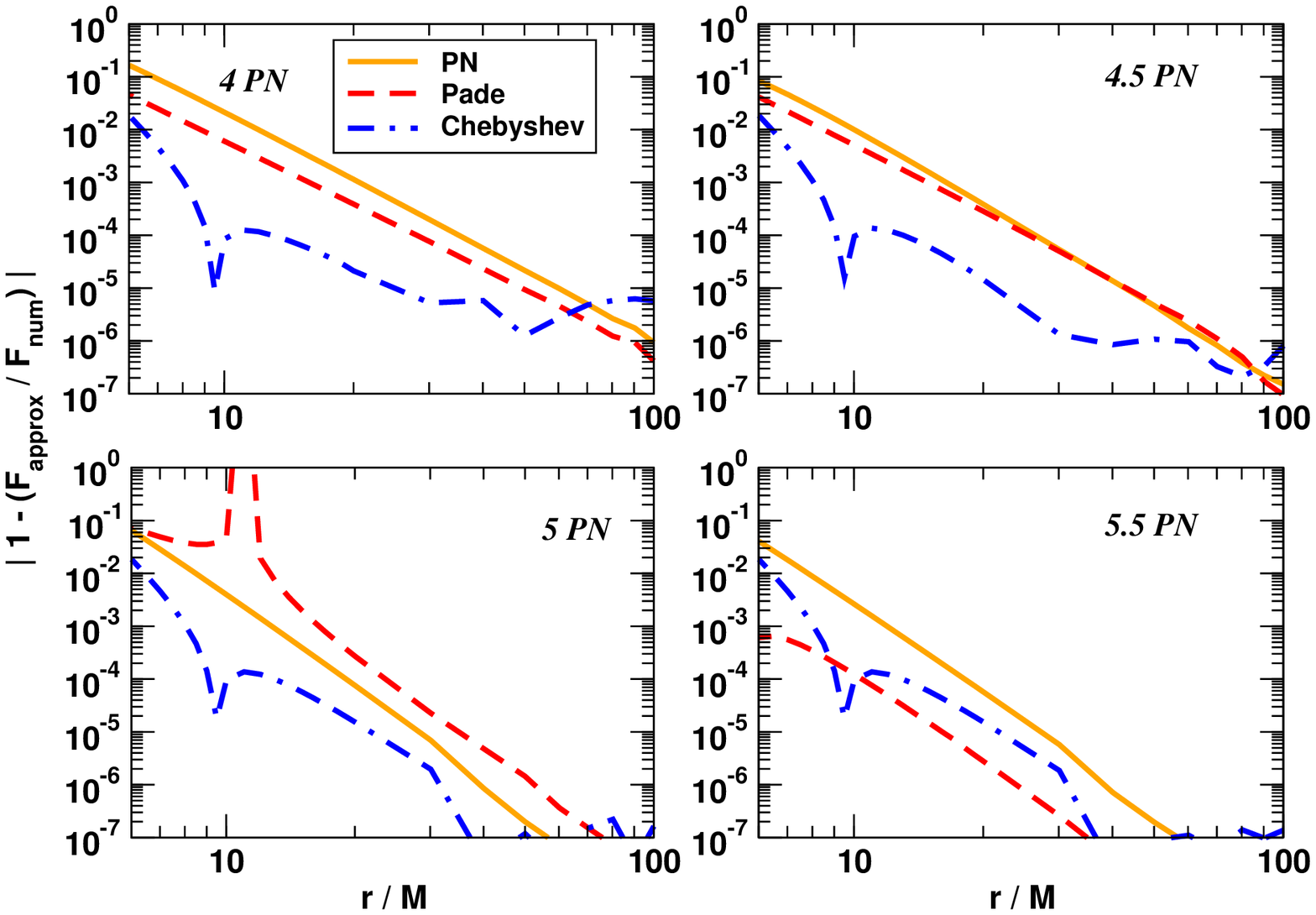}
    }
  }
\caption{The error of the PN, Pad\'e and Chebyshev approximations when compared to the numerical flux for 2 to 5.5 PN order. The flux is modelled from $R=100 M$ to the LSO at $R=6M$.}
\label{fig:cserror}
\end{figure}  
In Table~I we compare the induced truncation errors in both the PN and Chebyshev expansions, assuming that the maximum error in both cases is at the LSO.  In Figure~(\ref{fig:cserror}) we present the error of the PN, Pad\'e and Chebyshev approximations when compared to a numerical flux at 2 to 5.5 PN order.  We can see one of the downsides of the Pad\'e method in that the 2 PN approximation has a greater error at the LSO than the PN flux.  So in this case, the Pad\'e approximation really only begins to make a difference from 3 PN onwards.  We can see from this figure that the Chebyshev approximation tries to find an equal-ripple error curve.  It achieves this at high values of $r$, but as we approach the LSO the error begins to grow.  Over most of the plotted range, the Chebyshev approximations lie below both the PN and Pad\'e curves.  In order to do this, the Chebyshev approximation allows the error to grow in regions where we have good agreement in order to find a better fit elsewhere.  While the Chebyshev approximation is not as accurate as the PN or Pad\'e at high values of $r$, it has no influence on our results.  With the test-mass systems we have chosen, the lightest system, $(100,1.4)M_{\odot}$, comes into the detector bandwidth at about $15 M$, while the heaviest, $(50,1.4)M_{\odot}$, comes in at a greater value.  We can see from the graph that in the region $6 \leq  M \leq 40$, in nearly all cases, the Chebyshev approximation performs better than the PN and the Pad\'e at all orders of approximation. We should point out that the Chebyshev approximation is inferior to the Pad\'e approximation at the 5.5 PN level.  It is not understood why the Pad\'e flux behaves so well at this order.

\begin{figure}[t]
\vspace{0.25 in}
\begin{center}
\centerline{\epsfxsize=10cm \epsfysize=5cm \epsfbox{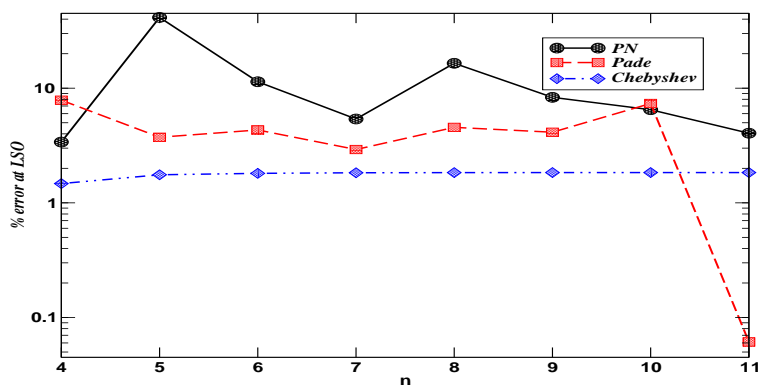}}
\caption{A logarithmic plot of the percentage error of the PN, Pad\'e and Chebyshev approximations to the gravitational wave flux when compared to the numerical flux at the LSO.}
\label{fig:errlso}
\end{center}
\end{figure}

We know that the main degradation in signal to noise ratio comes from a template going out of phase at the LSO.  In Figure~(\ref{fig:errlso}) we plot the percentage errors in each of the approximations at the LSO.  The PN approximation performs worst with errors of between $3.4\% \leq \epsilon_{_{PN}} \leq 42\%$, with the 2.5 PN approximation incurring the highest error.  Besides the 5.5 PN order where the error is $0.06\%$, the error for the Pad\'e flux at the LSO is $2.9\% \leq \epsilon_{_{Pade}} \leq 7.8\%$.  While we have plotted the Pad\'e error at 5 PN order, we must remember that there is a singularity at this order, so the result is superfluous.  In contrast, and a further sign that the Chebyshev approximation to the flux is related to the minimax polynomial, the error for the Chebyshev flux is approximately constant, $1.5\% \leq \epsilon_{_{Cheb}} \leq 1.8\%$, at all orders of approximation.  This is what we would expect as it attempts to minimize the maximum error at the LSO.

\section{Results and Discussion.}\label{sec:results}
We use the technique of matched filtering to see just how well each template performs.  For our `exact' signal we use a restricted PN waveform where we use the exact expression for the orbital energy function, and a numerical gravitational wave flux function.  We define the noise-weighted overlap between two waveforms $h(t)$ and $s(t)$ as the inner product of the normalized waveforms denoted by
\begin{equation}
{\mathcal O} = \frac{\left<h\left|s\right>\right.}{\sqrt{\left<h\left|h\right>\right.\left<s
\left|s\right>\right.}},
\end{equation}
where the scalar product is defined by 
\begin{equation}
\left<h\left|s\right.\right> =2\int_{0}^{\infty}\frac{df}{S_{h}(f)}\,\left[ \tilde{h}(f)\tilde{s}^{*}(f) +  \tilde{h}^{*}(f)\tilde{s}(f) \right].
\label{eq:scalarprod}
\end{equation}
Here, an asterix denotes a complex conjugate and a tilde denotes the Fourier transform of the time domain waveform.  Each template is a function of a number of parameters, $\lambda^{\mu}$, which in turn define the dimensionality of the search space.  For the types of systems we are considering, as well as the extrinsic parameters $t_{0}$ and $\phi_{0}$, each template is defined by the individual masses of the systems, $\left(m_{1},m_{2}\right)$.  Due to the short duration of the signal, parameters such as orbital inclination, position in the sky etc. are unimportant.  While we start off with a 4-d search space, $\lambda^{\mu} = \left\{t_{0},\phi_{0}, m_{1}, m_{2}\right\}$, we can reduce the search to the 2-d subspace of intrinsic parameters by maximizing over the extrinsic parameters.  We finally define the fitting factor $\mathcal{ FF}$ as the overlap maximized over all parameters
\begin{equation}
\mathcal{FF} = {\mathcal O}_{max_{\lambda^{\mu}}}.
\end{equation}
\begin{figure}
\vspace{0.25 in}
\centerline{\hbox{ \hspace{0.0in} 
    \epsfxsize=2.6in
    \epsffile{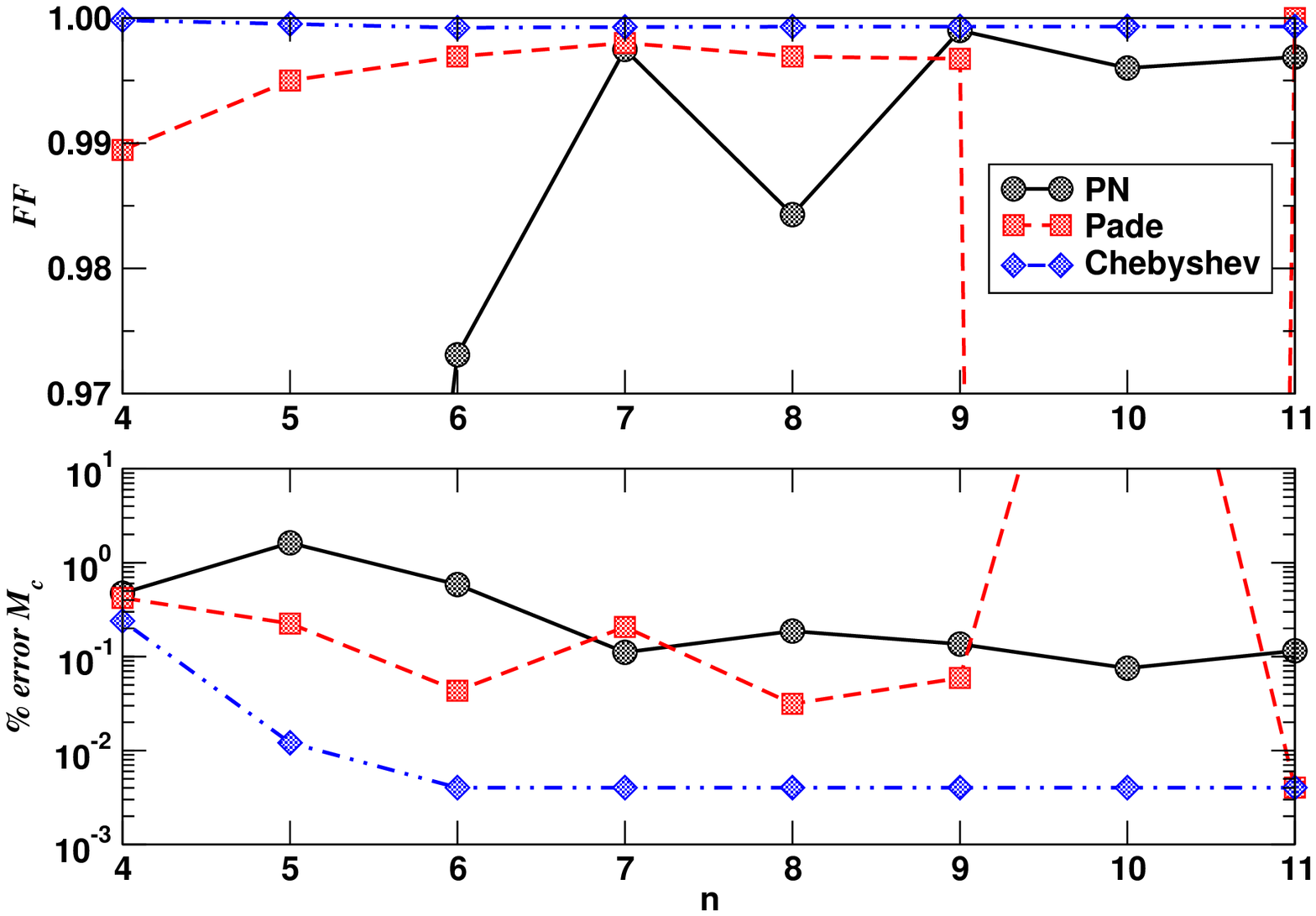}
    \hspace{0.25in}
    \epsfxsize=2.6in
    \epsffile{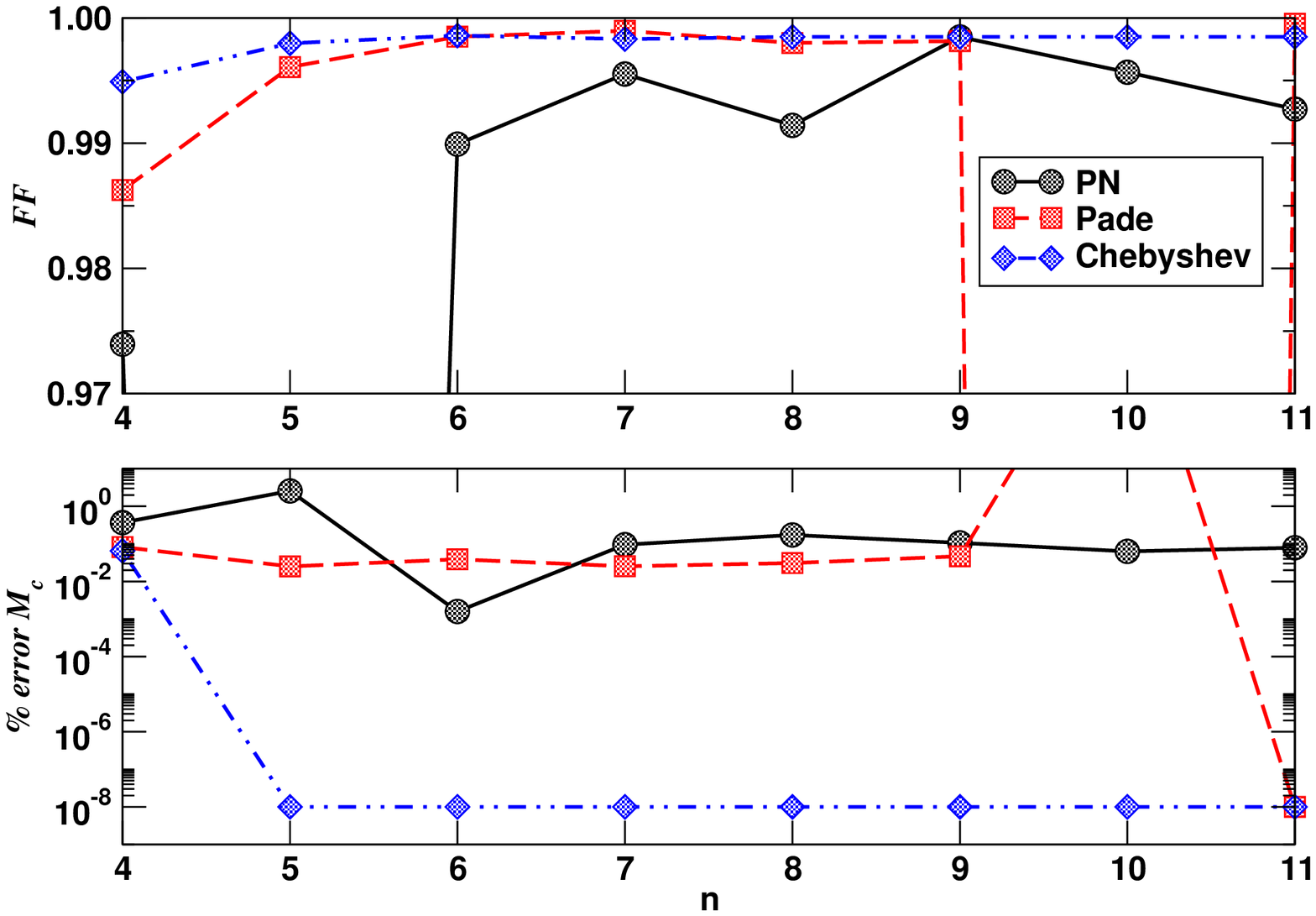}
    }
  }
\caption{The Fitting Factors (top) and \%-error in the estimation of the chirp mass (bottom) for PN, Pad\'e and Chebyshev templates as compared to an `exact' test-mass template with parameters of $(m_{1}, m_{2}) = (100, 1.4)M_{\odot}$ (left) and $(m_{1}, m_{2}) = (50, 1.4)M_{\odot}$ (right) .  The trough in the top cell and the peak in the bottom cell for the Pad\'e template at 5 PN is due to a singularity in the flux function at this order. }
\label{fig:TCME}
\end{figure}

For this study, as we are concentrating on the intermediate test-mass regime, we have chosen two systems to examine : $\left(100,1.4\right)$ and $\left(50,1.4\right)\,M_{\odot}$.  These correspond to reduced mass ratios of $\eta = 0.0136$ and  $0.0265$ respectively.  In order to calculate the fitting factors we used the expected EURO noise curve (as this gives us a lower frequency cutoff of about 10 Hz, thus giving us enough cycles to work with)~\cite{sathyapsd}.  While the waveforms are described by the two individual mass, it is impossible for a single ground based detector to detect these parameters.  What is possible is a combination of the individual mass called the chirp mass.  This is defined as $M_{c}=m\eta^{3/5}$.

On the left of Figure~(\ref{fig:TCME}) we plot the fitting factors and percentage errors in the estimation of the chirp mass.  The top cell shows the fitting factors for the PN, Pad\'e and Chebyshev templates for the masses $\left(100,1.4\right)M_{\odot}$, while the bottom cell displays the percentage error in the chirp-mass.  In general, we will take a template to be adequate if it reaches a predetermined threshold.  This is usually a fitting factor of about 0.97 .  In this case the PN templates at 2 and 2.5 PN achieve fitting factors of about 0.96 and 0.87 respectively making them inadequate as templates.  Only from 3 PN onwards do the templates meet the required threshold.  In keeping with everything that we know about the PN approximation, we can see that the fitting factors follow an oscillatory pattern, i.e. the 3.5 PN is better than both the 3 and 4 PN templates etc.  We can see from the bottom cell that not only do the lower order approximants achieve bad fitting factors, but they also have the highest errors in the estimation of $M_{c}$.  We can also see that the 4 PN template which performs worse than the 3.5 and 4.5 templates also has a higher parameter estimation error.  The Pad\'e templates achieve excellent fitting factors from 2 PN onwards, but again we can see an oscillatory nature in the estimation of the chirp mass.  While the 2 PN template has a fitting factor of $\sim 0.99$ it has an error in the estimation of $M_{c}$ almost equal to the PN template.   The error in the estimation of the chirp mass begins to improve as we increase the order of approximation, but we can see that the 3.5 template has a higher error in the chirp mass than the corresponding PN template.  We should also explain that the sudden dip in the fitting factor and peak in the chirp mass estimation at 5 PN is due to the singularity in the flux function giving a zero fitting factor and infinite error in the chirp mass.   At 5.5 PN order the Pad\'e template achieves almost a perfect overlap and the lowest error in $M_{c}$.   The Chebyshev templates, on the other hand, achieve fitting factors of almost unity at all orders of approximation.  Not only that, but we can see that the error in estimating $M_{c}$ improves, converging to a constant error value, which is again what we would expect as the template is near minimax.  Just looking at the error in parameter estimation, we can see that the Chebyshev templates have achieved by 3 PN order what takes the Pad\'e templates until 5.5 PN order to achieve.

On the right hand side of Figure~(\ref{fig:TCME}) we present the same results for the masses $\left(50,1.4\right)M_{\odot}$.  We can see in the top cell that while the 2 PN template now meets the require fitting factor, the 2.5 template again falls short.  Also, we notice that the oscillation in fitting factors is not as great as in the previous case with all PN templates achieving fitting factors of $>0.99$ from 3 PN order onwards.  Once again, the Pad\'e templates outperform the PN templates at all orders.  However, we again see that in order to do this, it means that the Pad\'e templates occasionally incur larger errors in the estimation of the chirp mass.  In this case we see that the 3 PN Pad\'e template has an error greater than the corresponding PN tempalate.  We also see this time that the error in the estimation of $M_{c}$ is only slightly better in a number of cases (2 PN, 3.5 PN, 4 PN and 4.5 PN orders) than the equivalent PN templates.   On the other hand, the Chebyshev templates again achieve fitting factors at all orders of approximation.  We also find, possibly due to the longer waveform and the accumulation of phase information, that while the error at 2 PN order is roughly an order of magnitude better than the PN template, from 2.5 PN order onwards the best-fit template has parameters almost identical to the signal we are trying to fit.  While hard to see from the plot, the error in the estimation of $M_{c}$ is approximately $10^{-8}\%$ from 2.5 to 5.5 PN order.  Once again, we can see that the 2.5 PN Chebyshev template performs as well as the 5.5 PN Pad\'e template and better than any of the PN templates.

\section{Conclusion.}
In this study we have introduced a new template for detecting gravitational waves from IMRI compact binary systems.  We have shown that one of the main problems with expansions such as the PN or Pad\'e approximations is that both adhere to the Weierstrass's theorem in that we need a large number of terms to properly approximate a function.  A better method is to use a member of the family of Ultraspherical polynomials, the Chebyshev polynomials of the first kind, which are closely related to the minimax polynomial due to the fact that they achieve equioscillation at $n+1$ points in an interval.  

We demonstrated that by using the shifted Chebyshev polynomials, we can define a new Chebyshev approximation to the gravitational wave flux function.  A major advantage of this new flux function is that when we expand a series in terms of Chebyshev polynomials, each subsequent coefficient is smaller than the previous one.  As the shifted polynomials have a maximum value of unity over the domain of interest, it means that the truncation error incurred by going to lower orders of approximation is proportional to the neglected coefficient, which is in general a small number, and is much smaller than the truncation error involved in the PN approximation.

By graphically fitting the PN, Pad\'e and Chebyshev approximations to the gravitational wave flux function to a numerical flux, we were able to show that at all orders of approximation the Chebyshev approximation provided a better fit than the PN approximation, and was better than the Pad\'e approximation at all orders except the 5.5 PN order.  The closeness of the Chebyshev flux to the minimax flux was observed due to the fact that the new flux tried to achieve an `equal-ripple' error curve.  One of the main results of this study is the fact the new flux function has a lower error at the LSO, where the templates are most likely to be out of phase with a signal, than both the PN and Pad\'e approximations.

Finally, by choosing a number of test-mass systems and a fiducially `exact' signal, we found that not only does the Chebyshev template always achieve higher fitting factors, but they have excellent error estimation, in most cases finding the parameters almost perfectly.  Another of the main features of the new templates are their ability to perform as well at lower orders of approximation than either the PN or Pad\'e templates at the highest order.  We believe that these new templates will prove to be an invaluable addition to any strategy that involves the detection of gravitational waves using both ground and eventually, space-based detectors.

\appendix

\section{Fitting Factors and Parameter Estimation}\label{app:ffs}
We present the values of the fitting factors and the error in the estimation of the Chirp Mass in the following tables.  The labels T, P and C stand for the PN, Pad\'e and Chebyshev results.  Row one of each order of approximation denotes the fitting factor, in row two are the individual masses associated with each fitting factor and row three corresponds to the percentage error in the estimation of the Chirp Mass.  The blank entries at 5 PN order for the Pad\'e templates is due to a singularity at this order of approximation in the gravitational wave flux function.

\begin{table}[h]
\begin{tabular}{|c|c c c|c c c|}\hline \hline
 & \multicolumn{3}{c|}{$(100, 1.4)M_{\odot}$} & \multicolumn{3}{c|}{$(50, 1.4)M_{\odot}$}    \\ \hline
$n$ & $T$ & $P$ & $C$ & $T$ & $P$ & $C$ \\ \hline\hline
4 & 0.9614 & 0.9895 & 0.9998 & 0.974 & 0.9863 & 0.9949   \\
 & $\left(98.83 , 1.4\right)$ & $\left(102.13 , 1.39\right)$ &  $\left(99.54 , 1.41\right)$ & $\left(49.03 , 1.41\right)$ & $\left(51.72 , 1.37\right)$ &  $\left(50.08 , 1.4\right)$  \\
 & 0.473 & 0.422 & 0.24 & 0.37 & 0.083 & 0.065    \\ \hline
5 & 0.866 & 0.995 & 0.9995 & 0.602 & 0.996 & 0.998    \\
 & $\left(116.19 , 1.3\right)$ & $\left(100.56 , 1.4\right)$ & $\left(100.03 ,1.4 \right)$ & $\left(56.15 , 1.35\right)$ & $\left(50.56 , 1.39\right)$ & $\left(50 ,1.4 \right)$   \\
 & 1.62 & 0.225 & 0.012 & 2.57 & 0.025 & 0  \\ \hline
6 & 0.973 & 0.9969 & 0.9992 & 0.9899 & 0.9985 & 0.9986   \\
 & $\left(97.52 ,1.41 \right)$ & $\left(100.96 , 1.39\right)$ &  $\left(100.01 ,1.4 \right)$ & $\left(47.98 ,1.44 \right)$ & $\left(50.48 , 1.39\right)$ &  $\left(49.99 ,1.4 \right)$   \\
 & 0.585 & 0.043 & 0.004  & 0.002 & 0.039 & 0   \\ \hline
7 & 0.9975 & 0.998 & 0.9993 & 0.9955 & 0.9989 & 0.9984      \\
 & $\left( 100.79,1.39 \right)$ & $\left(100.55 ,1.39 \right)$ &  $\left(100.01 ,1.4 \right)$ & $\left( 50.41,1.39 \right)$ & $\left(50.21 ,1.39 \right)$ &  $\left(50 ,1.4 \right)$   \\
 & 0.111 & 0.207 & 0.004 & 0.095 & 0.211 & 0 \\ \hline
8 & 0.9843 & 0.9969 &  0.9993 & 0.9914 & 0.9979 &  0.9984   \\
 & $\left(103.91 ,1.36 \right)$ & $\left(100.99 ,1.39 \right)$ &  $\left(100.01 ,1.4 \right)$ & $\left(51.95 ,1.36 \right)$ & $\left(50.49 ,1.39 \right)$ &  $\left(50 ,1.4 \right)$   \\
 & 0.187 & 0.031 &  0.004 & 0.173 & 0.031 &  0 \\ \hline
9 & 0.999 & 0.9967 & 0.9993 & 0.9984 & 0.9981 & 0.9985   \\
 & $\left(98.25 ,1.42 \right)$ & $\left(100.92 ,1.39 \right)$ &  $\left(100.01 ,1.4 \right)$ & $\left(49.1 ,1.42 \right)$ & $\left(50.47 ,1.39 \right)$ &  $\left(50 ,1.4 \right)$  \\
 & 0.136 & 0.059 &  0.004 & 0.107 & 0.047 &  0   \\ \hline
10 & 0.996 & -- &  0.9993 & 0.9957 & -- &  0.9985   \\
 & $\left(100.88 ,1.39 \right)$ & $\left(- , -\right)$ & $\left(100.01 ,1.4 \right)$ & $\left(50.45 ,1.39 \right)$ & $\left(- , -\right)$ & $\left(50 ,1.4 \right)$   \\
 & 0.075 & -- & 0.004 & 0.063 & -- & 0   \\ \hline
11 & 0.9969 & 0.9999 & 0.9993 & 0.9927 & 0.9995 & 0.9919  \\
 & $\left(100.78 ,1.39 \right)$ & $\left(100 ,1.4 \right)$ &  $\left(100.01 ,1.4 \right)$ & $\left(50.43 ,1.39 \right)$ & $\left(50 ,1.4 \right)$ &  $\left(50 ,1.4 \right)$   \\
 & 0.115 & 0.004 & 0.004 & 0.079 & 0 & 0    \\ \hline\hline
 
\end{tabular}
\end{table}

\end{document}